# On the Entropy Rate of Pattern Processes

George M. Gemelos    Tsachy Weissman

Department of Electrical Engineering
Stanford University
350 Serra Mall
Stanford, CA 94305, USA
Email: {ggemelos, tsachy}@stanford.edu

### Abstract

We study the entropy rate of pattern sequences of stochastic processes, and its relationship to the entropy rate of the original process. We give a complete characterization of this relationship for i.i.d. processes over arbitrary alphabets, stationary ergodic processes over discrete alphabets, and a broad family of stationary ergodic processes over uncountable alphabets. For cases where the entropy rate of the pattern process is infinite, we characterize the possible growth rate of the block entropy.

## 1 Introduction

In their recent work [11], Orlitsky et al. consider the compression of sequences with unknown alphabet size. This work, among others, has created interest in examining random processes with arbitrary alphabets which may a priori be unknown. One can think of this as a problem of reading a foreign language for the first time. As one begins to parse characters, one's knowledge of the alphabet grows. Since the characters in the alphabet have initially no meaning beyond the order in which they appear, one can relabel these characters by the order of their first appearance. Given a string, we refer to the relabeled string as the *pattern* associated with the original string.

**Example 1** *Assume that the following English sentence was being parsed into a pattern by a non-English speaker.*

$$\text{english is hard to learn}\ldots$$

*The associated pattern would be*

$$1, 2, 3, 4, 5, 6, 7, 8, 5, 6, 8, 7, 9, 10, 11, 8, 12, 13, 8, 4, 1, 9, 10, 2, \ldots$$

*regarding the space too as a character.*

We abstract this as follows: given a stochastic process $\mathbf{X} = \{X_i\}_{i \geq 1}$, we create a pattern process $\mathbf{Z} = \{Z_i\}_{i \geq 1}$.

It is the compression of the pattern process $\{Z_i\}$ that is the focus of both [1] and [11]. This emphasis is justified by the fact that the bulk of the information is in the pattern. Although universal compression is an extensively studied problem, the universal compression of pattern sequences is relatively new, see [6, 7, 8, 10, 11, 12, 13, 14, 15, 17]. The majority of these recent papers address universality questions of how well a pattern sequence associated with an unknown source can be compressed relative to the case where this distribution is known. Emphasis is on quantifying the redundancy, i.e., the difference between what can be achieved with and without knowledge of the source distribution. The main question we focus on in this work is how the entropy rate of a sequence and that of its pattern relate. More specifically, our goal is to study the relationship between the entropy rate $H(\mathbf{X})$ of the original process[1] $\{X_i\}$, and the entropy rate $H(\mathbf{Z})$ of the associated pattern process. This relationship is not always trivial, as the following examples illustrate.

**Example 2** *Let $X_i$ be drawn i.i.d. $\sim P$, where $P$ is a pmf on a finite alphabet. Then we show below that $H(\mathbf{X}) = H(\mathbf{Z})$.*

The intuition behind this result is that given enough time, all the symbols with positive probability will be seen, after which time the original process and its associated pattern sequence coincide, up to relabeling of the alphabet symbols.

**Example 3** *Let $X_i$ be drawn i.i.d. $\sim$ uniform $[0, 1]$. Then the entropy rate of $\{X_i\}$ is $\infty$. Since the probability of seeing the same value twice is zero, $Z_i = i$ w.p. 1 for all $i$ and, consequently, $H(\mathbf{Z}) = 0$.*

The connection between the entropy rate of the pattern and that of the original process was first studied for i.i.d. processes by Shamir and Song in [17]. The results in [17] give bounds on the block entropy of the pattern with respect to the block entropy of the original process. Such bounds naturally extend to bounds on the entropy rate. These bounds are improved upon in [14, 15, 16]. The work in [14, 15, 16, 17] is primarily focused on finite block entropy. Although such results are extremely useful for gaining insight into the finite block entropy behavior, a question different from the one we present here, they do not completely characterize the relationship between the entropy rate of an i.i.d. process and that of its associated pattern. The first complete characterization of this entropy rate relationship for the general i.i.d. case as well as Markov, noise-corrupted and finite alphabet stationary ergodic processes, is given in [3]. Orlitsky et al. in [10] independently derive the relationship for i.i.d. processes. The finite alphabet stationary ergodic result of [3] were later extended to general finite entropy discrete stationary ergodic processes in [4], and independently for finite entropy discrete ergodic processes in [9]. The uncountable alphabet i.i.d.

---

[1]Throughout this work, $X_m^n$ will denote the sequence $X_m, X_{m+1}, \ldots, X_n$. If not specified, $m$ will be assumed to be 1. Furthermore, $H(\mathbf{X})$ will denote entropy rate throughout this work, regardless of the discreteness of the distributions of $\{X^n\}$ (it should thus be regarded as $\infty$ when these are not discrete).

result of [3] were also extended to a family of uncountable alphabet processes with memory in [4] and [5]. The proof techniques used by Orlitsky et al. in [9, 10] are significantly different than those used in [3, 4], the latter we present here.

In this work we characterize the relationship between process and pattern entropy rates for general i.i.d., discrete Markov, and discrete stationary ergodic processes. Although the discrete Markov case falls under the more general results for discrete stationary ergodic sources, it will be shown that there is insight to be gained by exploring the discrete Markov case on its own. We then move on to examine stationary ergodic processes, with memory, over uncountable alphabets. In particular, we consider the Markov and additive noise case. These two results are then used to show a more general result for a broad family of stationary ergodic processes over uncountable alphabets. Finally, for the case where the entropy rate of the pattern process is infinite, we examine the possible growth rates for the block entropy of pattern processes.

In Section 2 we characterize the relationship between process and pattern entropy rate for the case of a generally distributed i.i.d. process. In Section 3, we examine the discrete Markov and the general discrete stationary ergodic process. Furthermore, in Section 4 we extend the uncountable alphabet results of Section 2 to certain processes with memory. In Section 5, we characterize a set of achievable asymptotic growth rates for the block entropy of a pattern process. We conclude in Section 6 with a brief summary of our results.

## 2 The I.I.D. Case

Consider the case where $X_i$ are generated i.i.d. $\sim f$, where $f$ is an arbitrary distribution on the arbitrary source alphabet $\mathcal{A}$. Let $S = \{x \in \mathcal{A} : \Pr\{X_1 = x\} > 0\}$.

**Theorem 1** *Given $X_i$ i.i.d. $\sim f$ and $\{Z_i\}$ its associated pattern process, for an arbitrary $x_o \notin S$ define the process*

$$\tilde{X}_i = \begin{cases} X_i & \text{if } X_i \in S \\ x_o & \text{otherwise.} \end{cases}$$

*Then*

$$H(\mathbf{Z}) = H(\tilde{\mathbf{X}}) = H(\tilde{X}_1),$$

*regardless of the finiteness of both sides of the equality.* [2]

Since we will make use of Corollary 6 in the proof of Theorem 1, we present the proof in Appendix A. It should be noted that Theorem 1 was independently discovered by Orlitsky et al. in [10]. As can be seen, Theorem 1 is consistent with Example 2 and Example 3. Note that the process $\{\tilde{X}_i\}$ is created by keeping all the point masses in $S$ and assigning all the remaining probability to a new point mass. This corresponds with the result in Example 3 which suggests that the pattern of a process drawn according to a pdf has no randomness, i.e. an entropy rate of zero. Therefore, the only randomness in the pattern comes from the point masses and the event of falling on a "non-point-mass-mode".

---

[2]Throughout this work, we use $H$ to denote both entropy rate, when the argument is a process, and entropy, when the argument is a random variable.

**Example 4** *Let $\{X_i\}$ be an i.i.d. process with each component drawn, with probability $1/3$, as $N(0,1)$ and, with probability $2/3$, as Bernoulli $1/2$. In this case $\tilde{X}_i$ is uniformly distributed on an alphabet of size 3. Therefore, Theorem 1 gives $H(\mathbf{Z}) = \log(3)$.*

Although $|S| < \infty$ in all three examples above, it should be noted that Theorem 1 makes no such assumption.

# 3 Discrete Alphabet Processes

Having characterized the relationship between process and pattern entropy rate for the general i.i.d. process, what can be said about processes with memory? To begin exploring the answer to this question we examine one of the most basic stationary ergodic processes with memory, the Markov process.

## A  Markov Processes over Discrete Alphabets

Although discrete Markov processes fall under the more general Theorem 3 to follow, which deals with discrete stationary ergodic processes, there is insight gained by examining the Markov case on it own. In particular, we will see that the proof of the general discrete stationary ergodic result relies heavily on a version of the Shannon-McMillan-Breiman theorem for countably infinite alphabets, found in [2], while no such heavy machinery is necessary for the simpler Markov case. This fact is due to the inherent structure of a Markov process and makes the Markov case an interesting example on it own. Later on in Section 4, we will also see it is this structure which makes the Markov process the first candidate for the extension of the uncountable alphabet results of Section 2 to uncountable alphabet processes with memory.

The entropy rate of Markov processes is well-known. What can be said about the entropy rate of the associated pattern processes? We first look at the case of a first order Markov process with components in a countable alphabet.

**Proposition 1** *Let $\{X_i\}$ be a stationary ergodic first order Markov process on the countable alphabet $\mathcal{A}$ and let $\{Z_i\}$ be the associated pattern process. If $H(\mathbf{X}) < \infty$, then*
$$H(\mathbf{Z}) = H(\mathbf{X}).$$

*Proof of Proposition 1:*
Let $\mu$ be the stationary distribution of the Markov process and let $P_x(y) = P(X_{t+1} = y | X_t = x)$ for all $x, y \in \mathcal{A}$. The data processing inequality implies $H(X^n) \geq H(Z^n)$ for all $n$. Hence
$$H(\mathbf{X}) \geq \limsup_{n \to \infty} \frac{1}{n} H(Z^n).$$

To complete the proof it remains to show

$$\liminf_{n \to \infty} \frac{1}{n} H(Z^n) \geq H(\mathbf{X}),$$

for which we will need the following three lemmas.

**Lemma 1** *If $\{x_n\}_{n=1}^{\infty}$ is a non-negative sequence, then*

$$\liminf_{n \to \infty} \frac{1}{n} \sum_{i=1}^{n} x_i \geq \liminf_{n \to \infty} x_n.$$

*Proof of Lemma 1:*
Since the sequence $\{x_n\}_{n=1}^{\infty}$ is non-negative, then

$$\frac{1}{n} \sum_{i=1}^{n} x_i \geq \frac{1}{n} \sum_{i=\lfloor \sqrt{n} \rfloor}^{n} x_i \quad \forall n \in \mathbb{N}.$$

Therefore,

$$\begin{aligned}
\liminf_{n \to \infty} \frac{1}{n} \sum_{i=1}^{n} x_i &\geq \liminf_{n \to \infty} \frac{1}{n} \sum_{i=\lfloor \sqrt{n} \rfloor}^{n} x_i \\
&\geq \liminf_{n \to \infty} \frac{n - \lfloor \sqrt{n} \rfloor + 1}{n} \inf \left\{ x_i : i \geq \lfloor \sqrt{n} \rfloor \right\} \\
&= \lim_{n \to \infty} \frac{n - \lfloor \sqrt{n} \rfloor + 1}{n} \inf \left\{ x_i : i \geq \lfloor \sqrt{n} \rfloor \right\} \\
&= \lim_{n \to \infty} \inf \left\{ x_i : i \geq \lfloor \sqrt{n} \rfloor \right\} \\
&= \liminf_{n \to \infty} x_n.
\end{aligned}$$

□

**Lemma 2** *Let $\{A_n\}$ and $\{B_n\}$ be two sequences of events such that $\lim_{n \to \infty} P(A_n) = 1$ and $\lim_{n \to \infty} P(B_n) = b$. Then $\lim_{n \to \infty} P(A_n \cap B_n) = b$.*

*Proof of Lemma 2:*
$P(A_n \cap B_n) \leq P(B_n) \to b$. On the other hand,

$$\begin{aligned}
\liminf_{n \to \infty} P(A_n \cap B_n) &= \liminf_{n \to \infty} 1 - P(A_n^c \cup B_n^c) \\
&\geq \liminf_{n \to \infty} 1 - P(A_n^c) - P(B_n^c) \\
&= 1 - 0 - (1 - b).
\end{aligned}$$

□

**Lemma 3** *Given any $B \subseteq \mathcal{A}$ such that $|B| < \infty$*

$$\liminf_{n \to \infty} \frac{1}{n} H(Z^n) \geq \sum_{b \in B} \mu(b) H(\Phi_B[P_b]),$$

*where we define the pmf $\Phi_B[P_b](x) = 1_B(x) P_b(x) + P_b(B^c) \delta_{x_o}(x)$, for an arbitrary $x_o \notin B$.* [3] *Here $\delta_{x_o}$ is used to denote the distribution which places unit mass on $x_o$.*

For an arbitrary distribution $P$ on alphabet $\mathcal{A}$ and $B \in \mathcal{A}$, $\Phi_B[P]$ can be thought of as the distribution created by keeping distribution $P$ on the set $B$ and clumping the remaining probability $P\{B^c\}$ on a single new point mass.

*Proof of Lemma 3:*
Let $A(X^n)$ be the set of distinct elements in $\{X_1, \ldots, X_n\}$. Then

$$\liminf_{n \to \infty} \frac{1}{n} H(Z^n) = \liminf_{n \to \infty} \frac{1}{n} \sum_{i=1}^{n} H(Z_i | Z^{i-1})$$

$$\stackrel{(a)}{\geq} \liminf_{n \to \infty} H(Z_n | Z^{n-1})$$

$$\stackrel{(b)}{\geq} \liminf_{n \to \infty} H(Z_n | X^{n-1})$$

$$\stackrel{(c)}{=} \liminf_{n \to \infty} H(Z_n | X_{n-1}, A(X^{n-1}))$$

$$\stackrel{(d)}{=} \liminf_{n \to \infty} H\left(Z_n | X_{n-1}, A(X^{n-1}), \mathbb{1}_{\{B \subseteq A(X^{n-1})\}}\right)$$

$$= \Pr\{B \not\subseteq A(X^{n-1})\} H\left(Z_n | X_n, A(X^{n-1}), \mathbb{1}_{\{B \subseteq A(X^{n-1})\}} = 0\right)$$
$$+ \Pr\{B \subseteq A(X^{n-1})\} H\left(Z_n | X_n, A(X^{n-1}), \mathbb{1}_{\{B \subseteq A(X^{n-1})\}} = 1\right)$$

$$\geq \Pr\{B \subseteq A(X^{n-1})\} H\left(Z_n | X_n, A(X^{n-1}), \mathbb{1}_{\{B \subseteq A(X^{n-1})\}} = 1\right)$$

$$\geq \sum_{b \in B} \Pr\{X_{n-1} = b, B \subseteq A(X^{n-1})\} H\left(Z_n | X_n = b, A(X^{n-1}), \mathbb{1}_{\{B \subseteq A(X^{n-1})\}} = 1\right)$$

$$\stackrel{(e)}{\geq} \sum_{b \in B} \Pr\{X_{n-1} = b, B \subseteq A(X^{n-1})\} H(Z_n | X_n = b, A(X^{n-1}) = B)$$

$$= \sum_{b \in B} \Pr\{X_{n-1} = b, B \subseteq A(X^{n-1})\} H(\Phi_B[P_b])$$

$$\stackrel{(f)}{\geq} \sum_{b \in B} \mu(b) H(\Phi_B[P_b])$$

where $(a)$ comes from Lemma 1, $(b)$ from the data processing inequality, and $(c)$ from the fact that Markovity implies that $Z_n$ is independent of $X^{n-2}$ given $X_{n-1}$ and $A(X^{n-1})$, (d) from the data processing inequality, (e) from a combination of Jensen's inequality and the data processing inequality, and (f) from Lemma 2 since, $\Pr\{B \subseteq A(X^n)\} \to 1$. □

---

[3]Throughout this work, given a distribution $f$ and a set $B$, $\Phi_B[f]$ will denote the distribution defined by $\Phi_B[f](x) = 1_B(x) f(x) + f(B^c) \delta_{x_o}(x)$ for an arbitrary $x_o \notin B$. When $f$ is a distribution, $H(f)$ will denote the entropy of a random variable drawn according to $f$. Furthermore, $1_A$ will denote the indicator function on the set $A$, while $\mathbb{1}_A$ will denote the indicator random variable on the event $A$.

Now let $\{B_k\}$ be a sequence of sets such that $B_k \subseteq \mathcal{A}$, $|B_k| < \infty$ for all $k$, and

$$\lim_{k \to \infty} \sum_{b \in B_k} \sum_{a \in B_k} -\mu(b) P_b(a) \log P_b(a) = \sum_{b \in \mathcal{A}} \sum_{a \in \mathcal{A}} -\mu(b) P_b(a) \log P_b(a),$$

regardless of the finiteness of both sides of the equation. Note that since the above summands are all positive, such a sequence $\{B_k\}$ can always be found. Lemma 3 gives

$$\liminf_{n \to \infty} \frac{1}{n} H(Z^n) \geq \sum_{b \in B_k} \mu(b) H(\Phi_{B_k}[P_b]) \quad \forall \ k.$$

Hence, by taking $k \to \infty$, we get

$$\liminf_{n \to \infty} \frac{1}{n} H(Z^n) \geq \lim_{k \to \infty} \sum_{b \in B_k} \mu(b) H(\Phi_{B_k}[P_b])$$
$$\geq \lim_{k \to \infty} \sum_{b \in B_k} \mu(b) \sum_{a \in B_k} -P_b(a) \log P_b(a)$$
$$= \lim_{k \to \infty} \sum_{b \in B_k} \sum_{a \in B_k} -\mu(b) P_b(a) \log P_b(a)$$
$$\stackrel{(a)}{=} \sum_{b \in \mathcal{A}} \sum_{a \in \mathcal{A}} -\mu(b) P_b(a) \log P_b(a)$$
$$\stackrel{(b)}{=} H(\mathbf{X}),$$

where (a) comes from the construction of $\{B_k\}$ and (b) from the fact that $\{X_i\}$ is a finite entropy first order Markov process. Note that (b) is not necessarily true for infinite entropy first order Markov processes. □

One should note that the proof of Proposition 1 can easily be extended to the case of Markov processes of any order. Hence, without going through the proof, we state the following:

**Theorem 2** *Let $\{X_i\}$ be a stationary ergodic Markov process of order $m$ on the countable alphabet $\mathcal{A}$, and let $\{Z_i\}$ be the associated pattern process. If $H(\mathbf{X}) < \infty$, then*

$$H(\mathbf{Z}) = H(\mathbf{X}).$$

## B  Stationary Ergodic Processes over Discrete Alphabets

Now that we have characterized the entropy rate relationship for the discrete Markov process, the natural next step would be to extend the results to all stationary ergodic processes on a countable alphabet.

**Theorem 3** *Let $\{X_i\}$ be a stationary ergodic process with components taking values from the countable alphabet $\mathcal{A}$, and assume $H(\mathbf{X}) < \infty$. Let $\{Z_i\}$ be the associated pattern process. Then*

$$H(\mathbf{Z}) = H(\mathbf{X}).$$

We will see that as compared to the proof of Proposition 1, the structure of the proof of Theorem 3 is slightly different, using a sandwich argument, and making use of heavier machinery such as a version of the Shannon-McMillan-Breiman theorem for countably infinite alphabets [2].

It is also important to note that like Theorem 2, Theorem 3 also has a finite entropy constraint. The need to exclude processes with infinite entropy from Theorem 3 is a direct result of the requirement of finite entropy for the countably infinite version of the Shannon-McMillan-Breiman theorem.

The proof of Theorem 3 will use the following two claims.

**Claim 1** *Let $(Z_0^{(n)}, \ldots, Z_{-n}^{(n)})$ denote the pattern of the sequence $(X_0, \ldots, X_{-n})$.*

$$\lim_{n \to \infty} H(Z_0^{(n)} | X_{-n}^{-1}) = H(X_0 | X_{-\infty}^{-1}).$$

*Proof of Claim 1*:
It is sufficient to show

$$\lim_{n \to \infty} H(X_0 | X_{-n}^{-1}) = H(X_0 | X_{-\infty}^{-1}) \tag{1}$$

and

$$\lim_{n \to \infty} |H(X_0 | X_{-n}^{-1}) - H(Z_0^{(n)} | X_{-n}^{-1})| = 0. \tag{2}$$

From [2] we know that $-\log(P(X_0 | X_{-n}^{-1})) \to -\log(P(X_0 | X_{-\infty}^{-1}))$ a.s. and the sequence is uniformly integrable, implying (1).

Moving on to (2) we see that the data processing inequality gives us $H(X_0 | X_{-n}^{-1}) \geq H(Z_0^{(n)} | X_{-n}^{-1})$ for all $n$. Hence it will suffice to show

$$\limsup_{n \to \infty} H(X_0 | X_{-n}^{-1}) - H(Z_0^{(n)} | X_{-n}^{-1}) \leq 0.$$

Let $A\left(X_{-n}^{-1}\right)$ be the set of distinct elements in $\{X_{-1}, \ldots, X_{-n}\}$. Then

$$H(X_0|X_{-n}^{-1}) - H(Z_0^{(n)}|X_{-n}^{-1}) = \mathbb{E}\left[\sum_{x \in \mathcal{A}} P_{X_0|X_{-n}^{-1}}(x) \log\left(\frac{1}{P_{X_0|X_{-n}^{-1}}(x)}\right)\right]$$

$$- \mathbb{E}\left[\sum_{x \in A\left(X_{-n}^{-1}\right)} P_{X_0|X_{-n}^{-1}}(x) \log\left(\frac{1}{P_{X_0|X_{-n}^{-1}}(x)}\right)\right]$$

$$- \mathbb{E}\left[P_{X_0|X_{-n}^{-1}}\left(A\left(X_{-n}^{-1}\right)^c\right) \log\left(\frac{1}{P_{X_0|X_{-n}^{-1}}\left(A\left(X_{-n}^{-1}\right)^c\right)}\right)\right]$$

$$= \mathbb{E}\left[\sum_{x \in A\left(X_{-n}^{-1}\right)^c} P_{X_0|X_{-n}^{-1}}(x) \log\left(\frac{1}{P_{X_0|X_{-n}^{-1}}(x)}\right)\right]$$

$$- \mathbb{E}\left[P_{X_0|X_{-n}^{-1}}\left(A\left(X_{-n}^{-1}\right)^c\right) \log\left(\frac{1}{P_{X_0|X_{-n}^{-1}}\left(A\left(X_{-n}^{-1}\right)^c\right)}\right)\right]$$

$$\leq \mathbb{E}\left[\sum_{x \in A\left(X_{-n}^{-1}\right)^c} P_{X_0|X_{-n}^{-1}}(x) \log\left(\frac{1}{P_{X_0|X_{-n}^{-1}}(x)}\right)\right]. \tag{3}$$

Since $H(X_0) < \infty$, given $\epsilon > 0$ there exists a $B \subset \mathcal{A}$ such that: $|B| < \infty$, if $b \in B$ then $\Pr\{X_0 = b\} > 0$, and

$$H(\Phi_{B^c}[P_X]) \leq \epsilon, \tag{4}$$

where $P_X$ is the distribution on $X_0$ and $\Phi_{B^c}[P_X]$ is defined as in Lemma 3. Since

$$\mathbb{E}\left[\sum_{x \in B^c} P_{X_0|X_{-n}^{-1}}(x) \log\left(\frac{1}{P_{X_0|X_{-n}^{-1}}(x)}\right)\right] \leq H(\Phi_{B^c}[P_\mathbf{X}]|X_{-n}^{-1}) \leq H(\Phi_{B^c}[P_\mathbf{X}]) \quad \forall n,$$

(4) implies

$$\mathbb{E}\left[\sum_{x \in B^c} P_{X_0|X_{-n}^{-1}}(x) \log\left(\frac{1}{P_{X_0|X_{-n}^{-1}}(x)}\right)\right] \leq \epsilon \quad \forall n. \tag{5}$$

By the ergodicity of $\{X_i\}$

$$\lim_{n \to \infty} \Pr\{B \subset A\left(X_{-n}^{-1}\right)\} = 1. \tag{6}$$

From (3) and the construction of $B$ we get

$$H(X_0|X_{-n}^{-1}) - H(Z_0^{(n)}|X_{-n}^{-1}) \leq \mathbb{E}\left[\mathbb{1}_{\{B \subset A(X_{-n}^{-1})\}} \sum_{x \in A(X_{-n}^{-1})^c} P_{X_0|X_{-n}^{-1}}(x) \log\left(\frac{1}{P_{X_0|X_{-n}^{-1}}(x)}\right)\right]$$

$$+ \mathbb{E}\left[\mathbb{1}_{\{B \not\subset A(X_{-n}^{-1})\}} \sum_{x \in A(X_{-n}^{-1})^c} P_{X_0|X_{-n}^{-1}}(x) \log\left(\frac{1}{P_{X_0|X_{-n}^{-1}}(x)}\right)\right]$$

$$\leq \mathbb{E}\left[\mathbb{1}_{\{B \subset A(X_{-n}^{-1})\}} \sum_{x \in B^c} P_{X_0|X_{-n}^{-1}}(x) \log\left(\frac{1}{P_{X_0|X_{-n}^{-1}}(x)}\right)\right]$$

$$+ H(X_0|X_{-n}^{-1}) \mathbb{E}\left[\mathbb{1}_{\{B \not\subset A(X_{-n}^{-1})\}}\right]$$

$$\leq \mathbb{E}\left[\mathbb{1}_{\{B \subset A(X_{-n}^{-1})\}} \sum_{x \in B^c} P_{X_0|X_{-n}^{-1}}(x) \log\left(\frac{1}{P_{X_0|X_{-n}^{-1}}(x)}\right)\right]$$

$$+ H(X_0) \mathbb{E}\left[\mathbb{1}_{\{B \not\subset A(X_{-n}^{-1})\}}\right]$$

$$\stackrel{(a)}{\leq} \epsilon \mathbb{E}\left[\mathbb{1}_{\{B \subset A(X_{-n}^{-1})\}}\right] + H(X_0) \mathbb{E}\left[\mathbb{1}_{\{B \not\subset A(X_{-n}^{-1})\}}\right]$$

$$\leq \epsilon \Pr\{B \subset A(X_{-n}^{-1})\} + H(X_0)\left(1 - \Pr\{B \subset A(X_{-n}^{-1})\}\right),$$

where $(a)$ follows from (5). Taking the limit in $n$, (6) gives

$$\limsup_{n \to \infty} H(X_0|X_{-n}^{-1}) - H(Z_0^{(n)}|X_{-n}^{-1}) \leq \epsilon.$$

Since $\epsilon$ was arbitrary, (2) follows, completing the proof of Claim 1. □

**Claim 2**
$$H(\mathbf{X}) = H(X_0|X_{-\infty}^{-1}).$$

*Proof of Claim 2:*
From [2] we know that $-\log(P(X_0|X_{-n}^{-1})) \to -\log(P(X_0|X_{-\infty}^{-1}))$ a.s. and the sequence is uniformly integrable. Therefore, uniform integrability and almost sure convergence implies convergence in mean. □

We are now ready for:

*Proof of Theorem 3:*

$$H(\mathbf{X}) = \limsup_{n \to \infty} \frac{1}{n} H(X^n)$$
$$\stackrel{(a)}{\geq} \limsup_{n \to \infty} \frac{1}{n} H(Z^n)$$
$$\geq \liminf_{n \to \infty} \frac{1}{n} H(Z^n)$$

$$= \liminf_{n\to\infty} \frac{1}{n}\sum_{i=0}^{n-1} H(Z_{i+1}|Z^i)$$

$$\stackrel{(b)}{\geq} \liminf_{n\to\infty} H(Z_{n+1}|Z^n)$$

$$\stackrel{(c)}{=} \liminf_{n\to\infty} H\left(Z_0^{(n)} \,\Big|\, Z_{-1}^{(n)}, \ldots, Z_{-n}^{(n)}\right)$$

$$\geq \liminf_{n\to\infty} H(Z_0^{(n)}|X_{-n}^{-1})$$

$$\stackrel{(d)}{=} H(X_0|X_{-\infty}^{-1})$$

$$\stackrel{(e)}{=} H(\mathbf{X}),$$

where $(a)$ follows from the data processing inequality, $(b)$ from Lemma 1, $(c)$ is a result of stationarity, $(d)$ results from Claim 1, and $(e)$ results from Claim 2. As a reminder, we use $(Z_0^{(n)}, \ldots, Z_{-n}^{(n)})$ to denote the pattern of the sequence $(X_0, \ldots, X_{-n})$. □

## 4 Uncountable Alphabet Processes with Memory

The i.i.d. results of Theorem 1 completely characterize the entropy rate relationship for the general memoryless stationary process. So far, we have only addressed the case of discrete processes with memory. A natural question that arises is whether the relationship between the entropy rate of the process and that of the pattern shown in Theorem 1 can be extended to processes with memory over an uncountable alphabet?

Besides helping to answer the question of how far we can extend the i.i.d. results of Theorem 1, the study of the uncountable alphabet setting is also motivated by real world processes such as discrete signals which are jittered. Any discrete process corrupted by Gaussian noise can be thought of as an example of such jittered processes. Although the motivation of lossless compression is not as applicable in the uncountable alphabet setting, patterns may still be useful. In general, focusing on the pattern allows us to map our process into a finite alphabet process. Although information is lost in the mapping, the pattern may still capture relevant information and therefore prove to be useful in certain applications such as lossy compression.

Furthermore, the study of continuous alphabets allows us to look at the effect of densities on the entropy relationship. Although densities are strictly a property of continuous alphabets, they can be used to better understand the finite block behavior of the entropy relationship in the discrete setting. In particular, when looking at a finite block length $n$, it is possible for a discrete process to have a subset of the support which has large measure, but whose elements each have measure much smaller than $1/n$. Taking the limit in $n$, no such set can exist for discrete processes, but for finite $n$ such a set acts like an effective density and affects the entropy relationship for finite blocks. An example of the role of such an effective density can be found in [16] where bounds on the finite block entropy of patterns generated by i.i.d. processes are developed. In [16], Shamir concludes the paper with the observation that low probability symbols contribute to the pattern entropy mostly as a single super-symbol, which is exactly how Theorem 1 describes the contribution of the density part of a distribution to the entropy rate of patterns generated by i.i.d. processes. Hence the study of the continuous alphabet setting may not only extend the limit results of Theorem 1, but also give insight into the finite block behavior of the entropy relation for the general discrete setting. With this motivation in mind, we begin our examination of the entropy rate relationship in the uncountable alphabet setting by first looking at Markov processes.

## A  Markov Processes over Uncountable Alphabets

We observed in Section 3 that the inherent structure of the Markov process simplified the proof of the results in the discrete case. The hope is that by looking at this heavily structured family first we will develop some insight into the more general case of a stationary ergodic process over an uncountable alphabet.

Although we are unable to characterize the entropy rate of the induced pattern process for a general uncountable alphabet Markov process, the following proposition covers a fairly general family of Markov processes. Before we state the proposition, let us generalize some of the notation used in Proposition 1. Given an $m^{th}$ order Markov process $\{X_i\}$ on $\mathbb{R}$, for $x^m \in \mathbb{R}^m$ let $f_{x^m}$ be the kernel associated with the state $x^m$. We will denote the set of point masses of $f_{x^m}$ as $S_{x^m}$, $S_{x^m} = \{y \in \mathbb{R} : \Pr\{X_{m+1} = y | X^m = x^m\} > 0\}$.

**Proposition 2** *Let $\{X_i\}$ be a stationary ergodic Markov process on $\mathbb{R}$ of order $m$ such that there exists $S \subset \mathbb{R}$ with $S_{x^m} = S$ for all $x^m \in \mathbb{R}^m$ and $\Phi_S[f_{x^m}] = \Phi_S[f_{y^m}]$ for all $x^m, y^m \notin S^m$. Let $\{Z_i\}$ be the pattern process associated with $\{X_i\}$. Define the process $\{\tilde{X}_i\}$ as*

$$\tilde{X}_i = \begin{cases} X_i & \text{if } X_i \in S \\ x_o & \text{otherwise} \end{cases}$$

*for an arbitrary $x_o \in S^c$. If $|S| < \infty$, then*

$$H(\mathbf{Z}) = H(\tilde{\mathbf{X}}).$$

The proof of Proposition 2, as well as the remaining results of the present section begins with the observation that Theorem 3 implies $H(\tilde{\mathbf{X}}) = H(\tilde{\mathbf{Z}})$, where $\{\tilde{Z}_i\}$ is the pattern process associated with $\{\tilde{X}_i\}$. It is then left to show that $H(\mathbf{Z})$ is equal to $H(\tilde{\mathbf{Z}})$. To this end, we show that for any given $n$, the difference between $H(Z^n)$ and $H(\tilde{Z}^n)$ is either bounded or grows sub-linearly in $n$.

*Proof of Proposition 2:*
If $|S| = 0$, w.p. 1 the process $\{X_i\}$ does not repeat and therefore $H(\mathbf{Z}) = 0$. Similarly if $|S| = 0$, the process $\{\tilde{X}_i\}$ is a constant and therefore $H(\tilde{\mathbf{X}}) = 0$. Hence $H(\mathbf{Z}) = H(\tilde{\mathbf{X}})$.

We now look at the case where $|S| > 0$. We observe that since $\Phi_S[f_{x^m}] = \Phi_S[f_{y^m}]$ for all $x^m, y^m \notin S^m$, $\{\tilde{X}_i\}$ is a discrete Markov process of order $m$. Hence Theorem 2 and the fact that $\{\tilde{X}_i\}$ is stationary ergodic and has finite entropy gives

$$H(\tilde{\mathbf{X}}) = H(\tilde{\mathbf{Z}}). \tag{7}$$

For $x \in S$ define the waiting time $I_x = \inf\{i > 0 : X_i = x\}$. Given $\{I_x\}_{x \in S}$ we know the first appearance of every point in $S$. Hence we know the first appearance of every point but those which are assigned zero probability by every kernel, i.e. all but those that appear at most once w.p. 1. Therefore given $\tilde{Z}^n$ and $\{I_x\}_{x \in S}$ we can reconstruct $Z^n$ w.p. 1 for all $n$. Similarly given $Z^n$ and $\{I_x\}_{x \in S}$ we can reconstruct $\tilde{Z}^n$ for all $n$. Hence

$$H(Z^n) \le H(\tilde{Z}^n) + \sum_{x \in S} H(I_x) \quad \forall n \tag{8}$$

and

$$H(Z^n) + \sum_{x \in S} H(I_x) \ge H(\tilde{Z}^n) \quad \forall n. \tag{9}$$

**Claim 3**
$$H(I_x) < \infty \quad \forall x \in S$$

*Proof of Claim 3:*
Given $x \in S$, define $d_{\min} = \min\{\Pr\{X_{m+1} = x | X^m = y^m\} : y^m \in \mathbb{R}^m\}$ and $d_{\max} = \max\{\Pr\{X_{m+1} = x | X^m = y^m\} : y^m \in \mathbb{R}^m\}$. Note that since $0 < |S| < \infty$ and $\Phi_S[f_{x^m}] = \Phi_S[f_{y^m}]$ for all $x^m, y^m \notin S^m$, $d_{\min}$ and $d_{\max}$ are well defined. By the definition of $S$, $d_{\min} > 0$.

First, let us consider the case where $d_{\max} = 1$. Then, there exists a $x \in S$ and $y^m \in \mathbb{R}^m$ such that $\Pr\{X_{m+1} = x | X^m = y^m\} = 1$ and $S = S_{y^m} = \{x\}$. We will first look at the case where $y^m \in S^m$. Since $S = \{x\}$, if $y^m \in S^m$, then $y^m = x^m$, where $x^m$ is the vector $(x, \ldots, x)$ of length $m$. Therefore $f_{x^m} = \delta_x$ and once the state $x^m$ is reached it cannot be exited. Hence in order for $\{\tilde{X}_i\}$ to be irreducible, which is required for the process to be ergodic, it must place zero or unit probability on being in state $x^m$. By the construction of $S$ and the fact that $x \in S$, $\Pr\{X^m = x^m\} > 0$ and therefore $\Pr\{\tilde{X}^m = x^m\} > 0$. Hence $X_i = x$ w.p. 1 and $H(I_x) = 0$. Let us now examine the case where $y^m \notin S^m$. Since $\Phi_S[f_{u^m}] = \Phi_S[f_{y^m}]$ for all $u^m, y^m \notin S^m$, if $u^m \notin S^m$, then $\Pr\{X_{m+1} = x | X^m = u^m\} = 1$. Noting that $S = \{x\}$, we can conclude that if $X_i \ne x$, then w.p. 1 $X_{i+1} = x$. Hence $I_x \le 2$ and $H(I_x) \le \log 2$.

We now consider the less trivial case where $d_{\max} < 1$. Since regardless of the state $y^m$, $\Pr\{X_{m+1} = x | X^m = y^m\} \in [d_{\min}, d_{\max}]$, then

$$\Pr\{I_x = i\} \in [d_{\min}(1 - d_{\max})^{i-1}, d_{\max}(1 - d_{\min})^{i-1}] \quad \forall i \in \mathbb{N}.$$

Hence

$$H(I_x) = \sum_{i=1}^{\infty} \Pr\{I_x = i\} \log\left(\frac{1}{\Pr\{I_x = i\}}\right)$$

$$\leq \sum_{i=1}^{\infty} d_{\max}(1-d_{\min})^{i-1} \log\left(\frac{1}{d_{\min}(1-d_{\max})^{i-1}}\right)$$

$$= -d_{\max} \sum_{i=1}^{\infty} (1-d_{\min})^{i-1} \left[\log(d_{\min}) + \log((1-d_{\max})^{i-1})\right]$$

$$= -d_{\max} \sum_{i=1}^{\infty} (1-d_{\min})^{i-1} \log(d_{\min})$$

$$\quad - d_{\max} \sum_{i=1}^{\infty} (i-1)(1-d_{\min})^{i-1} \log(1-d_{\max})$$

$$= -d_{\max} \log(d_{\min}) \sum_{i=0}^{\infty} (1-d_{\min})^{i}$$

$$\quad - d_{\max} \log(1-d_{\max}) \sum_{i=0}^{\infty} i(1-d_{\min})^{i}$$

$$= -\frac{d_{\max} \log(d_{\min})}{d_{\min}} - \frac{d_{\max}(1-d_{\min})\log(1-d_{\max})}{d_{\min}^2}. \tag{10}$$

Since $d_{\min}, d_{\max} \in (0,1)$, equation (10) implies $H(I_x) < \infty$. □

Claim 3 therefore gives

$$\lim_{n \to \infty} \frac{H(I_x)}{n} = 0 \quad \forall x \in S. \tag{11}$$

Combining equations (8), (9), (11) and noting that $|S| < \infty$ gives $H(\mathbf{Z}) = H(\tilde{\mathbf{Z}})$. Equation (7) then completes the proof. □

**Example 5** *Let $\{X_i\}$ be a first order Markov process on $[0,1]$ with the following transition kernels, represented as generalized densities on $[0,1]$:*

$$f_0(y) = \frac{3}{4}\delta_0(y) + \frac{1}{4}\delta_1(y)$$

$$f_1(y) = \frac{1}{4}\delta_0(y) + \frac{1}{2}\delta_1(y) + \frac{1}{4},$$

*and for $x \in (0,1)$*

$$f_x(y) = \frac{1}{4}\delta_0(y) + \frac{1}{4}\delta_1(y) + \frac{3}{4}\mathbf{1}_{\{(x-1/2)(y-1/2)>0\}}(y)$$

$$\quad + \frac{1}{4}\mathbf{1}_{\{(x-1/2)(y-1/2)\leq 0\}}(y).$$

*It is readily checked that the stationary distribution given the above kernels is*

$$\mu(x) = \frac{1}{2}\delta_0(x) + \frac{1}{3}\delta_1(x) + \frac{1}{6} \quad \forall x \in [0,1].$$

In the above case, $\{\tilde{X}_i\}$ can be thought of as a first order Markov process on the set $\{0, 1/2, 1\}$, (the value $1/2$ chosen arbitrarily) with transition probabilities whose generalized densities are

$$\tilde{f}_0(y) = \frac{3}{4}\delta_0(y) + \frac{1}{4}\delta_1(y)$$
$$\tilde{f}_1(y) = \frac{1}{4}\delta_0(y) + \frac{1}{2}\delta_1(y) + \frac{1}{4}\delta_{1/2}(y)$$
$$\tilde{f}_{1/2}(y) = \frac{1}{4}\delta_0(y) + \frac{1}{4}\delta_1(y) + \frac{1}{2}\delta_{1/2}(y).$$

Hence $\{\tilde{X}_i\}$ has the following stationary distribution

$$\tilde{\mu}(x) = \frac{1}{2}\delta_0(x) + \frac{1}{3}\delta_1(x) + \frac{1}{6}\delta_{1/2}(x).$$

Applying Proposition 2 gives

$$H(\mathbf{Z}) = H(\tilde{\mathbf{X}})$$
$$= \frac{1}{2}H(\tilde{X}_2|\tilde{X}_1 = 0) + \frac{1}{3}H(\tilde{X}_2|\tilde{X}_1 = 1) + \frac{1}{6}H(\tilde{X}_2|\tilde{X}_1 = 1/2)$$
$$= \frac{1}{2}\left(2 - \frac{3}{4}\log 3\right) + \frac{1}{3}\left(\frac{3}{2}\right) + \frac{1}{6}\left(\frac{3}{2}\right)$$
$$= \frac{7}{4} - \frac{3}{8}\log(3) = 1.1556.$$

## B  Additive White Noise-Corrupted Processes

We now consider the case of a noise-corrupted process. Let $\{X_i\}$ be a stationary ergodic process and $\{Y_i\}$ be its noise-corrupted version. Here we assume i.i.d. additive noise, $\{N_i\}$, with $X_i$, $Y_i$, and $N_i$ taking values in $\mathbb{R}$. Let $S_Y$ and $S_N$ denote the set of point masses for $Y_i$ and $N_i$ respectively. We will also define the process

$$\tilde{N}_i = \begin{cases} N_i & \text{if } N_i \in S_N \\ n_o - X_i & \text{otherwise} \end{cases}$$

for an arbitrary $n_o \notin S_Y$.

**Proposition 3** *Let $\{X_i\}$ be a finite alphabet stationary ergodic process. Let $\{Y_i\}$ and $\{\tilde{Y}_i\}$ denote the process $\{X_i\}$ corrupted by the additive noise $\{N_i\}$, and $\{\tilde{N}_i\}$, respectively. Further let $\{Z_i\}$ denote the pattern process associated with $\{Y_i\}$. If $|S_N| < \infty$, then*

$$H(\mathbf{Z}) = H(\tilde{\mathbf{Y}}).$$

It is interesting to note that the result of Proposition 3 can be rephrased to look more like those of Theorem 1 and Proposition 2. This is accomplished by observing that the process $\{\tilde{Y}_i\}$, used in Proposition 3, can also be constructed by

$$\tilde{Y}_i = \begin{cases} Y_i & \text{if } Y_i \in S_Y \\ n_o & \text{otherwise} \end{cases}$$

for an arbitrary $n_o \notin S_Y$. This is the construction of $\{\tilde{X}_i\}$ used in both Theorem 1 and Proposition 3.

*Proof of Proposition 3:*
If $P(N_i \in S_N) = 1$ then $N_i = \tilde{N}_i$ and there is nothing to prove, so we will assume that $P(N_i \in S_N) < 1$. Let $\{\tilde{Z}_i\}$ denote the pattern process associated with the process $\{\tilde{Y}_i\}$. Since $\{\tilde{Y}_i\}$ is a discrete stationary ergodic process with finite entropy, Theorem 3 gives
$$H(\tilde{\mathbf{Z}}) = H(\tilde{\mathbf{Y}}).$$
Hence to complete the proof of Proposition 3, we just need to show that
$$H(\mathbf{Z}) = H(\tilde{\mathbf{Z}}).$$

Define
$$\hat{Z}(n)_i = \begin{cases} Z_i & \text{if } \exists\, j \in [1,n]\setminus i \text{ s.t. } Z_i = Z_j \\ y_o & \text{otherwise} \end{cases}$$
for some arbitrary non-integer $y_o$. Clearly $\hat{Z}(n)$ uniquely determines $Z^n$ and vice versa, so in particular,
$$H(Z^n) = H(\hat{Z}(n)) \quad \forall\, n > 0. \tag{12}$$

Define
$$I_{n_o} = \inf\{i > 0 : N_i \in S_N^c\}.$$
We also observe the following: if $\tilde{Z}_i \neq \tilde{Z}_{I_{n_o}}$ then $Y_i = \tilde{Y}_i$ and if $\tilde{Z}_i = \tilde{Z}_{I_{n_o}}$ then w.p. 1 $Y_i \neq Y_j$ for all $j \neq i$. Hence we can construct $\hat{Z}(n)$ from $\tilde{Z}^n$ and $I_{n_o}$ w.p. 1. Therefore
$$H(\tilde{Z}^n, I_{n_o}) \geq H(\hat{Z}(n)) \quad \forall\, n > 0$$
and consequently,
$$H(\tilde{Z}^n) + H(I_{n_o}) \geq H(\hat{Z}(n)) \quad \forall\, n > 0. \tag{13}$$

Since $I_{n_o}$ is the waiting time for the first appearance of an element from $S_N^c$ in the i.i.d. process $\{N_i\}$, it is geometrically distributed, and in particular has finite entropy. Therefore
$$\lim_{n \to \infty} \frac{H(I_{n_o})}{n} = 0$$
which combined with (12) and (13) gives
$$H(\tilde{\mathbf{Z}}) \geq H(\mathbf{Z}). \tag{14}$$

Defining $C(n)_i = \mathbb{1}_{\{\{\hat{Z}(n)_i = y_o\} \cap \{Y_i \in S_Y\}\}}$, we make the following observations: w.p. 1 $Y_i = \tilde{Y}_i$ if and only if $C(n)_i = 1$ or $\hat{Z}(n)_i \neq y_o$ and $\tilde{Y}_i = n_o$ if and only if $C(n)_i = 0$ and $\hat{Z}(n)_i = y_o$. From these observations we conclude that given $C(n)$ and $\hat{Z}(n)$ we can reconstruct $\tilde{Z}^n$ w.p. 1 for all $n > 0$. Hence, for all $n > 0$,

$$\begin{aligned} H(\tilde{Z}^n) &\leq H(\hat{Z}(n), C(n)) \\ &\leq H(\hat{Z}(n)) + H(C(n)) \\ &\stackrel{(a)}{=} H(Z^n) + H(C(n)) \\ &\leq H(Z^n) + \sum_{i=1}^{n} H(C(n)_i), \end{aligned} \quad (15)$$

where $(a)$ comes from (12).

Let
$$Pe_i^{(n)} = \Pr\{Y_i \in S_Y,\ Y_j \neq Y_i\ \forall\ j \in [1,n]\backslash i\}.$$

Then
$$\begin{aligned} Pe_i^{(n)} &= \Pr\{Y_i \in S_Y\} \Pr\{Y_j \neq Y_i\ \forall\ j \in [1,n]\backslash i\ |Y_i \in S_Y\} \\ &= \Pr\{Y_i \in S_Y\} \sum_{y \in S_Y} \Pr\{Y_j \neq y\ \forall\ j \in [1,n]\backslash i\ |Y_i = y\} \Pr\{Y_i = y\ |Y_i \in S_Y\} \\ &\leq \Pr\{Y_i \in S_Y\} \sum_{y \in S_Y} \Pr\{Y_j \neq y\ \forall\ j \in [1,n]\backslash i\ |Y_i = y\}. \end{aligned}$$

Without loss of generality assume that $i < n/2$

$$\begin{aligned} Pe_i^{(n)} &\leq \Pr\{Y_i \in S_Y\} \sum_{y \in S_Y} \Pr\{Y_j \neq y\ \forall\ j \in [i+1, i+n/2-1]|Y_i = y\} \\ &\stackrel{(a)}{\leq} \Pr\{Y_1 \in S_Y\} \sum_{y \in S_Y} \Pr\{Y_j \neq y\ \forall\ j \in [2, n/2]|Y_1 = y\}, \end{aligned}$$

where $(a)$ follows from the stationarity of $Y$. Let

$$Pe^{(n)} = \Pr\{Y_1 \in S_Y\} \sum_{y \in S_Y} \Pr\{Y_j \neq y\ \forall\ j \in [2, n/2]|Y_1 = y\}. \quad (16)$$

Therefore we have
$$Pe^{(n)} \geq Pe_i^{(n)} \quad \forall i. \quad (17)$$

By ergodicity we have
$$\lim_{n \to \infty} \Pr\{Y_j \neq y\ \forall\ j \in [2, n/2]|Y_1 = y\} = 0 \quad \forall y \in S_Y$$

and since $|S_Y| < \infty$, (16) gives us $\lim_{n \to \infty} Pe^{(n)} = 0$. Hence there exists an $N$ such that $Pe^{(n)} < 1/2$ for all $n > N$ and (17) implies that

$$H_B(Pe_i^{(n)}) \leq H_B(Pe^{(n)}) \quad \forall n \geq N, \quad (18)$$

where $H_B$ is the binary entropy function. Substituting $Pe_i^{(n)}$ into (15) and noting that $\mathbb{E}\left(C(n)_i\right) = Pe_i^{(n)}$ gives

$$\frac{H(\tilde{Z}^n)}{n} \leq \frac{H(Z^n)}{n} + \frac{1}{n}\sum_{i=1}^n H_B(Pe_i^{(n)}) \quad \forall n \geq 0$$

which combined with (18) gives us

$$\frac{H(\tilde{Z}^n)}{n} \leq \frac{H(Z^n)}{n} + H_B(Pe^{(n)}) \quad \forall n \geq N.$$

Since $\lim_{n\to\infty} Pe^{(n)} = 0$,

$$H(\tilde{\mathbf{Z}}) \leq H(\mathbf{Z}). \tag{19}$$

Combining (14) and (19) completes the proof. □

Note that in the case where $\{N_i\}$ is a discrete i.i.d. process, Proposition 3 agrees with Theorem 3. In the case where $N_i$ has no point masses then, as Example 3 would suggest, Proposition 3 gives $H(\mathbf{Z}) = 0$.

Having verified that Proposition 3 is in agreement with previous results, let us examine a case where previous theorems do not apply.

**Example 6** Let $\{X_i\}$ be a first order Markov process on the set $\{1, 2\}$ and let $\{N_i\}$ be i.i.d., independent of $\{X_i\}$, distributed according to the density

$$f_N(m) = \frac{1}{\sqrt{8\pi}}e^{-m^2/2} + \frac{1}{2}\delta_0(m),$$

where $\delta_0$ denotes a unit mass on 0. Further let $Y_i = X_i + N_i$ and $\{Z_i\}$ be its associated pattern process. Since $\{Y_i\}$ is a hidden Markov process with memory on a continuous alphabet, previous results fail to capture the relationship between $H(\mathbf{Y})$ and $H(\mathbf{Z})$. However, Proposition 3 gives

$$H(\mathbf{Z}) = H(\tilde{\mathbf{Y}}), \tag{20}$$

where $\tilde{Y}_i$ is the ternary hidden Markov process given by $X_i$ with probability $1/2$ and an arbitrary $n_o \notin \{1, 2\}$ with probability $1/2$. We can also use Proposition 3 to lower bound $H(\mathbf{Z})$ in terms of $H(\mathbf{X})$. Noting that $\{\tilde{Y}_i\}$ is simply $\{X_i\}$ with erasures, we let $I_i$ denote the event of erasure at time $i$. Then

$$\begin{aligned}
H(\tilde{Y}_n|\tilde{Y}^{n-1}) &\stackrel{(a)}{=} H(\tilde{Y}_n, I_n|\tilde{Y}^{n-1}) \\
&= H(I_n) + H(\tilde{Y}_n|\tilde{Y}^{n-1}, I_n) \\
&\stackrel{(b)}{=} H(I_n) + \Pr\{I_n = 0\}H(X_n|\tilde{Y}^{n-1}, I_n = 0) + 0 \\
&\stackrel{(c)}{\geq} H(I_n) + \Pr\{I_n = 0\}H(X_n|X^{n-1}, I_n = 0) \\
&\stackrel{(d)}{=} 1 + \frac{1}{2}H(X_n|X^{n-1})
\end{aligned} \tag{21}$$

where (a) follows from the fact that given $\tilde{Y}_i$ we know $I_i$, (b) from the fact that given $I_i = 1$, $\tilde{Y}_i$ is a constant and given $I_i = 0$, $\tilde{Y}_i = X_i$, (c) from a combination of the fact that $X_n$ is independent of $I^n$ and that conditioning decreases entropy, and finally, (d) follows from the fact that $\{I_i\}$ is an i.i.d. Bernoulli $1/2$ process, independent of the process $\{X_i\}$. Combining (20) and (21) we get

$$H(\mathbf{Z}) \geq \frac{1}{2} H(\mathbf{X}) + 1. \tag{22}$$

Note that (22) holds with equality when $\{X_i\}$ is i.i.d., as is readily seen to be implied by Theorem 1.

## C  Stationary Ergodic Processes over Uncountable Alphabets

Through the results of Proposition 2 and Proposition 3 we have seen two separate families of processes with memory on uncountable alphabets that share similar entropy rate properties. However, we are not able to extend such a relationship to the general stationary ergodic process. An interesting question that arises is what characteristics do the Markov processes of Proposition 2 and the additive noise processes of Proposition 3 share that allow for this characterization of the relationship between process and pattern entropy rates? In order to help answer this question, we examine the following Markov example which does not satisfy the requirements of Proposition 2.

**Example 7** *Let $\{X_i\}$ be a first order Markov process on $[0,1]$ with a uniform stationary distribution. Furthermore, conditioned on $X_i$, $X_{i+1} = X_i$ with probability $1/2$ and $X_{i+1}$ is drawn uniformly on $[0,1]$ with probability $1/2$. It is easy to see that $\{X_i\}$ does not satisfy the conditions of Proposition 2. In this case, $S = \{x \in [0,1] : \Pr\{X_1 = x\} > 0\} = \emptyset$ and therefore the sequence $\{\tilde{X}_i\}$ is constant and*

$$H(\tilde{\mathbf{X}}) = 0.$$

*We also observe that at any time $i+1$ we either see a new symbol with probability $1/2$ or we repeat $X_i$ with probability $1/2$. Therefore,*

$$H(\mathbf{Z}) = 1,$$

*not $H(\tilde{\mathbf{X}})$ as would be assumed from the relationship between pattern and process entropy rates found in Proposition 2 and Proposition 3. Hence, unlike the processes described in Proposition 2 and Proposition 3, we see that*

$$H(\mathbf{Z}) \neq H(\tilde{\mathbf{X}}).$$

Example 7 suggests that one of the important characteristics shared by the processes in Proposition 2 and Proposition 3, which allow for the equality between $H(\mathbf{Z})$ and $H(\tilde{\mathbf{X}})$, is the control over the repeatability of density points. In other words, assuring that for the most part only elements in $S$ are likely to be seen more than once. This characteristic is also demonstrated by the i.i.d. processes of Theorem 1, which share the equality between $H(\mathbf{Z})$ and $H(\tilde{\mathbf{X}})$.

With this in mind we can try to extend this characteristic to general stationary ergodic processes in hopes of developing a similar entropy rate relation. Before we state the next theorem, let us define some notation and make rigorous the criterion of repeatability described above. Given a stationary ergodic process $\{X_i\}$ on $\mathbb{R}$, let $S = \{x \in \mathbb{R} : \Pr\{X_1 = x\} > 0\}$ and $R = \{x \in \mathbb{R} : \Pr\{\exists j \geq 2 : X_j = X_1 | X_1 = x\} > 0\}$. Let $S = \{s_1, s_2, \ldots, s_{|S|}\}$ and $P_i = \Pr\{X_1 = s_i\}$. Without loss of generality we will assume that the elements of $S$ are ordered such that $P_1 \geq P_2 \geq \ldots$.

**Theorem 4** *Let $\{X_i\}$ be a stationary ergodic process on $\mathbb{R}$ with $\Pr\{X_1 \in R\} = \Pr\{X_1 \in S\}$. Let $\{Z_i\}$ be the associated pattern process. Define the process*

$$\tilde{X}_i = \begin{cases} X_i & \text{if } X_i \in S \\ x_o & \text{otherwise} \end{cases}$$

*for some arbitrary $x_o \notin S$. If $|S| < \infty$, then*

$$H(\mathbf{Z}) = H(\tilde{\mathbf{X}}).$$

*Otherwise, if $|S|$ is infinite and there exists $\beta > 2$ such that*

$$\lim_{i \to \infty} \frac{P_i}{(1/i)^\beta} = 0,$$

*then*

$$H(\mathbf{Z}) = H(\tilde{\mathbf{X}}).$$

It should be to noted that both Proposition 2 and Proposition 3 are special cases of Theorem 4.

The requirement $\Pr\{X_1 \in R\} = \Pr\{X_1 \in S\}$, is the mathematical equivalent of the statement that only elements in $S$ are likely to be seen more than once. While the $\beta$-convergence requirement is a technicality needed in the proof, it may prove to be non-essential.

Hence we see that controlling repeatability of density points is, essentially, a sufficient condition for establishing equality between $H(\mathbf{Z})$ and $H(\tilde{\mathbf{X}})$. Furthermore, Example 7 suggests that it is a necessary condition. Hence, the $\beta$-convergence requirement aside, there is reason to believe that Theorem 4 in some sense describes the largest family of stationary ergodic processes over uncountable alphabets for which the equality between $H(\mathbf{Z})$ and $H(\tilde{\mathbf{X}})$ holds. In particular, the $\beta$-convergence condition aside, Theorem 4 contains as special cases the i.i.d. results of Theorem 1, the discrete setting results of Theorem 3, the Markov results of Proposition 2, and the noise-corrupted process results of Proposition 3.

The proof of Theorem 4 begins with the observation that Theorem 3 can be used to show that $H(\tilde{\mathbf{X}}) = H(\tilde{\mathbf{Z}})$. We are then left to show that $H(\mathbf{Z}) = H(\tilde{\mathbf{Z}})$. This is done in a two step process. We first show that by making us of the information contained in the indexes of first appearance for a finite set $B \subseteq S$, we can bound the difference between $H(Z^n)$ and $H(\tilde{Z}^n)$. This bound is a function of $\Pr\{X_1 \in B\}$, $|B|$, and $n$ and is true for all $n$ and $B \subseteq S$. Finally. the limit condition on $\{P_i\}_{i=1}^{\infty}$ allows us to pick a sequence $\{B_n\}_{n=1}^{\infty}$ for which the upper bound on the difference between $H(Z^n)$ and $H(\tilde{Z}^n)$ grows sub-linearly in $n$, completing the proof of Theorem 4.

*Proof of Theorem 4:*
We can assume that $\Pr\{X_1 \in S\} < 1$, otherwise there is nothing to prove. If $|S| < \infty$, then $H(\tilde{X}_1) < \infty$. In the case where $|S|$ is infinite, then the fact that $\beta > 2$ and

$$\lim_{i \to \infty} \frac{P_i}{(1/i)^{\beta}} = 0,$$

implies that $H(\tilde{X}_1) < \infty$. Since $\{\tilde{X}_i\}$ is a discrete stationary ergodic process with finite entropy, Theorem 3 gives

$$H(\tilde{\mathbf{X}}) = H(\tilde{\mathbf{Z}}), \tag{23}$$

where $\{\tilde{Z}_i\}$ is the associated pattern process. To complete the proof of Theorem 4, we need to show that

$$H(\mathbf{Z}) = H(\tilde{\mathbf{Z}}).$$

For $x \in S$ define $I_x = \inf\{i > 0 : X_i = x\}$ and $I_x^{(n)} = I_x \mathbb{1}_{\{I_x \leq n\}} - \mathbb{1}_{\{I_x > n\}}$. Hence $I_x^{(n)}$ has an alphabet of size $n+1$ and therefore

$$H(I_x^{(n)}) \leq \log(n+1). \tag{24}$$

Given $B \subseteq S$ such that $|B| < \infty$, let $P_B = \Pr\{X_1 \in S \cap B^c\}$ and $C_B^{(n)} = \mathbb{1}_{\{X_1, X_2, \ldots, X_n \notin S \cap B^c\}}$.

If $C_B^{(n)} = 1$ then given $\{I_x^{(n)}\}_{x \in B}$, we know all the labels of the elements of $S$ which appear in $X^n$. Hence for $n \geq 1$ and conditioned on $C_B^{(n)} = 1$, given $Z^n$ and $\{I_x^{(n)}\}_{x \in B}$ we can reconstruct $\tilde{Z}^n$. Therefore

$$\begin{aligned}
\Pr\{C_B^{(n)} = 1\} H(\tilde{Z}^n | C_B^{(n)} = 1) &\leq \Pr\{C_B^{(n)} = 1\} H(\tilde{Z}^n, \{I_x^{(n)}\}_{x \in B} | C_B^{(n)} = 1) \\
&\leq \Pr\{C_B^{(n)} = 1\} H(Z^n, \{I_x^{(n)}\}_{x \in B} | C_B^{(n)} = 1) \\
&\leq \Pr\{C_B^{(n)} = 1\} H(Z^n | C_B^{(n)} = 1) \\
&\quad + \Pr\{C_B^{(n)} = 1\} H(\{I_x^{(n)}\}_{x \in B} | C_B^{(n)} = 1) \\
&\leq \Pr\{C_B^{(n)} = 1\} H(Z^n | C_B^{(n)} = 1) \\
&\quad + \Pr\{C_B^{(n)} = 0\} H(Z^n | C_B^{(n)} = 0) \\
&\quad + \Pr\{C_B^{(n)} = 1\} H(\{I_x^{(n)}\}_{x \in B} | C_B^{(n)} = 1) \\
&\quad + \Pr\{C_B^{(n)} = 0\} H(\{I_x^{(n)}\}_{x \in B} | C_B^{(n)} = 0) \\
&= H(Z^n | C_B^n) + H(\{I_x^{(n)}\}_{x \in B} | C_B^n) \\
&\leq H(Z^n) + H(\{I_x^{(n)}\}_{x \in B}) \\
&\leq H(Z^n) + |B| \log(n+1) \quad \forall n \geq 1. \tag{25}
\end{aligned}$$

Given $i$, we now wish to examine the probability

$$\Pr\{\exists j > i : X_j = X_i | X_i \notin S\} \stackrel{(a)}{=} \Pr\{\exists j \geq 2 : X_j = X_1 | X_1 \notin S\}$$
$$\stackrel{(b)}{\leq} \Pr\{\exists j \geq 2 : X_j = X_1 | X_1 \notin R\}$$
$$\leq \sum_{j=2}^{\infty} \Pr\{X_j = X_1 | X_1 \notin R\}, \quad (26)$$

where $(a)$ follows from the stationarity of the process $\{X_i\}$ and $(b)$ from the fact that $S \subseteq R$ and $\Pr\{X_1 \in R\} = \Pr\{X_1 \in S\}$ implies that $\Pr\{X_1 \in R \cap S^c\} = 0$.

To further bound $\Pr\{\exists j > i : X_j = X_i | X_i \notin S\}$ we now examine $\Pr\{X_j = X_1 | X_1 \notin R\}$. Let $f_j(x_1, x_j)$ be the measure on $(X_1, X_j)$ given $X_1 \notin R$. Therefore

$$\Pr\{X_j = X_1 | X_1 \notin R\} = \int_{x_1 \in R^c} \int_{x_j = x_1} f_j(x_1, x_j) dx_j dx_1. \quad (27)$$

Assume that $\Pr\{X_j = X_1 | X_1 \notin R\} > 0$, then (27) implies that there exists $x_1 \in R^c$ such that

$$\int_{x_j = x_1} f_j(x_1, x_j) dx_j > 0.$$

This is only possible if

$$\Pr\{X_1 = X_j = x_1 | X_1 \notin R\} > 0. \quad (28)$$

Therefore,

$$\Pr\{X_j = X_1 | X_1 = x_1\} \stackrel{(a)}{=} \Pr\{X_j = X_1 | X_1 = x_1, x_1 \notin R\}$$
$$\geq \Pr\{X_j = X_1 | X_1 = x_1, x_1 \in R^c\} \Pr\{X_1 = x_1 | x_1 \notin R\}$$
$$= \Pr\{X_j = X_1 = x_1 | x_1 \notin R\}$$
$$\stackrel{(b)}{>} 0, \quad (29)$$

where $(a)$ follows from the fact that $x_1 \in R^c$ and $(b)$ from (28). By definition of $R$, (29) implies that $x_1 \in R$. This is a contradiction since $x_1 \in R^c$. Hence

$$\Pr\{X_j = X_1 | X_1 \notin R\} = 0 \quad \forall j \geq 2$$

and (26) gives

$$\Pr\{\exists j > i : X_j = X_i | X_i \notin S\} = 0. \quad (30)$$

Therefore w.p. 1, only elements in $S$ will appear more than once. Hence conditioned on $C_B^{(n)} = 1$, given $\{I_x^{(n)}\}_{x \in B}$ we know the labels of all the elements in $X^n$ except those that appear at most once w.p. 1. Therefore for $n \geq 1$ and conditioned on $C_B^{(n)} = 1$, given $\{I_x^{(n)}\}_{x \in B}$ and $\tilde{Z}^n$, we can reconstruct $Z^n$ w.p. 1. Hence

$$\begin{aligned}
\Pr\{C_B^{(n)} = 1\}H(Z^n|C_B^{(n)} = 1) &\leq \Pr\{C_B^{(n)} = 1\}H(Z^n, \{I_x^{(n)}\}_{x \in B}|C_B^{(n)} = 1) \\
&\leq \Pr\{C_B^{(n)} = 1\}H(\tilde{Z}^n, \{I_x^{(n)}\}_{x \in B}|C_B^{(n)} = 1) \\
&\leq \Pr\{C_B^{(n)} = 1\}H(\tilde{Z}^n|C_B^{(n)} = 1) \\
&\quad + \Pr\{C_B^{(n)} = 1\}H(\{I_x^{(n)}\}_{x \in B}|C_B^{(n)} = 1) \\
&\leq \Pr\{C_B^{(n)} = 1\}H(\tilde{Z}^n|C_B^{(n)} = 1) \\
&\quad + \Pr\{C_B^{(n)} = 0\}H(\tilde{Z}^n|C_B^{(n)} = 0) \\
&\quad + \Pr\{C_B^{(n)} = 1\}H(\{I_x^{(n)}\}_{x \in B}|C_B^{(n)} = 1) \\
&\quad + \Pr\{C_B^{(n)} = 0\}H(\{I_x^{(n)}\}_{x \in B}|C_B^{(n)} = 0) \\
&= H(Z^n|C_B^n) + H(\{I_x^{(n)}\}_{x \in B}|C_B^n) \\
&\leq H(\tilde{Z}^n) + H(\{I_x^{(n)}\}_{x \in B}) \\
&\leq H(\tilde{Z}^n) + |B|\log(n+1) \quad \forall n \geq 1. \qquad (31)
\end{aligned}$$

If $|S| < \infty$, then set $B = S$. Therefore $P_B = 0$, $C_B^{(n)} = 1$ w.p. 1 and equations (25) and (31) give

$$\begin{aligned}
H(Z^n) &\leq H(\tilde{Z}^n) + |S|\log(n+1) \quad \forall n \geq 1, \\
H(\tilde{Z}^n) &\leq H(Z^n) + |S|\log(n+1) \quad \forall n \geq 1.
\end{aligned}$$

The finiteness of $|S|$ then implies

$$H(\mathbf{Z}) = H(\tilde{\mathbf{Z}}).$$

To complete the proof of Theorem 4 we need to address the case where $|S|$ is infinite. Choose $\gamma \in (0,1)$ such that $\alpha \triangleq \gamma(\beta - 1) > 1$. Such a $\gamma$ can be found since $\beta > 2$. Let $m(n) = \lceil n^\gamma \rceil$. Since

$$\lim_{i \to \infty} \frac{P_i}{(1/i)^\beta} = 0$$

and $m(n)$ is an unbounded increasing sequence, there exists $N > 0$ such that

$$m(n)^\beta P_{m(n)} < 1 \quad \forall n > N.$$

Construct a sequence of sets $B_n \subseteq S$ as follows, $B_n = \{s_1, \ldots, s_{m(n)}\}$ for all $n > N$. Therefore

$$\begin{aligned}
P_{B_n} &= \Pr\{X_1 \in S \cap B_n^c\} \\
&\leq \sum_{i=m(n)+1}^{\infty} \Pr\{X_1 = s_i\}
\end{aligned}$$

$$
\begin{aligned}
&= \sum_{i=m(n)+1}^{\infty} P_i \\
&\stackrel{(a)}{\leq} \sum_{i=m(n)+1}^{\infty} \frac{1}{i^\beta} \\
&\leq \int_{m(n)}^{\infty} \frac{1}{x^\beta} dx \\
&= \frac{1}{\beta-1}(m(n))^{1-\beta} \\
&= \frac{1}{\beta-1}\lceil n^\gamma \rceil^{1-\beta} \\
&\stackrel{(b)}{\leq} \frac{1}{\beta-1} n^{\gamma(1-\beta)} \\
&= \frac{1}{\beta-1} n^{-\alpha} \quad \forall n > N,
\end{aligned}
\qquad (32)
$$

where $(a)$ follows from the fact that $n > N$ and $(b)$ from the fact that $\beta > 2$. Therefore

$$
\begin{aligned}
H(\tilde{Z}^n) &\leq H(\tilde{Z}^n, C_{B_n}^{(n)}) \\
&= H(C_{B_n}^{(n)}) + H(\tilde{Z}^n | C_{B_n}^{(n)}) \\
&\leq 1 + \Pr\{C_{B_n}^{(n)} = 1\} H(\tilde{Z}^n | C_{B_n}^{(n)} = 1) + \Pr\{C_{B_n}^{(n)} = 0\} H(\tilde{Z}^n | C_{B_n}^{(n)} = 0) \\
&\leq 1 + \Pr\{C_B^{(n)} = 1\} H(\tilde{Z}^n | C_{B_n}^{(n)} = 1) + n P_{B_n} H(\tilde{Z}^n | C_{B_n}^{(n)} = 0) \\
&\leq 1 + \Pr\{C_B^{(n)} = 1\} H(\tilde{Z}^n | C_{B_n}^{(n)} = 1) + n P_{B_n} H(\tilde{Z}^n) \\
&\stackrel{(a)}{\leq} 1 + \Pr\{C_B^{(n)} = 1\} H(\tilde{Z}^n | C_{B_n}^{(n)} = 1) + n^2 P_{B_n} \log n \\
&\stackrel{(b)}{\leq} 1 + H(Z^n) + |B_n| \log(n+1) + n^2 P_{B_n} \log n \\
&= 1 + H(Z^n) + m(n) \log(n+1) + n^2 P_{B_n} \log n \\
&\stackrel{(c)}{\leq} 1 + H(Z^n) + (n^\gamma + 1) \log(n+1) + n^{2-\alpha} \log n \quad \forall n > N,
\end{aligned}
\qquad (33)
$$

where $(a)$ follows from the fact that $\tilde{Z}_i$ has an alphabet of at most $i$, $(b)$ follows from (25), and $(c)$ from (32). Similarly using (31) we get

$$
H(Z^n) \leq 1 + H(\tilde{Z}^n) + (n^\gamma + 1)\log(n+1) + n^{2-\alpha} \log n \quad \forall n > N. \qquad (34)
$$

Since $\gamma \in (0,1)$ and $\alpha > 1$, (33) and (34) give

$$
H(\mathbf{Z}) = H(\tilde{\mathbf{Z}}),
$$

completing the proof of Theorem 4. $\qquad\square$

# 5  Growth Rates

Now that we have explored the relationship between the entropy rate of the original process and the associated pattern process, we turn our attention to possible growth rates for the block entropy of a pattern sequence. In other words, having looked at the limit, we now look at the asymptotic growth rates.

**Theorem 5** *For any $\delta > 0$ there exists an i.i.d. process $\{X_i\}$ such that its associated pattern sequence satisfies*
$$\lim_{n \to \infty} \frac{H(Z_n|Z^{n-1})}{(\log n)^{1-\delta}} = \infty. \tag{35}$$

Note that since $Z_{n+1}$ lies in an alphabet of size at most $n+1$ we have, for any process, not even necessarily stationary,
$$\limsup_{n \to \infty} \frac{H(Z_n|Z^{n-1})}{\log n} \leq 1.$$

Theorem 5 then says that the growth rate $\log n$ is essentially, up to a factor which is sub-polynomial in $\log n$, achievable by an i.i.d. process. It should also be noted that the bounds on the block entropy of patterns generated by i.i.d. processed found in [14, 15, 16, 17] can be used to examine possible asymptotic growth rates for the entropy of pattern processes. An example of such an application can be found in [14].

We dedicate the remainder of this section to the proof of Theorem 5. Let $X_i$ be i.i.d. $\sim f_X$, where $X_i$ takes values in an arbitrary space $\mathcal{A}$, and $\{Z_i\}$ be the associated pattern sequence. Define $S = \{x \in \mathcal{A} : \Pr\{X_1 = x\} > 0\}$.

**Claim 4** $H(\Phi_B[f_X])$ *is increasing in $B$, i.e., for any $B_1 \subseteq B_2 \subseteq S$*
$$H(\Phi_{B_1}[f_X]) \leq H(\Phi_{B_2}[f_X]).$$

*Proof of Claim 4:*
This is nothing but a data-processing inequality. Indeed, let $Y \sim \Phi_{B_2}[f_X]$ and let
$$U = \begin{cases} Y & \text{if } Y \in B_1 \\ x_o & \text{otherwise.} \end{cases}$$

Clearly $U \sim \Phi_{B_1}[f_X]$ and $U$ is a deterministic function of $Y$, thus the claim follows. $\square$

**Proposition 4** *For any $B \subseteq S$*
$$H(Z_{n+1}|Z^n) \geq H(\Phi_B[f_X]) \left[1 - |B| \exp\left(-n \min_{b \in B} \Pr\{X = b\}\right)\right].$$

*Proof of Proposition 4:*
Letting $P_X^n$ denote the distribution of $X^n$, for any $B \subseteq S$,
$$H(Z_{n+1}|Z^n) \geq H(Z_{n+1}|X^n)$$

$$\begin{aligned}
&= \int_{\mathcal{A}^n} H(Z_{n+1}|X^n = x^n) dP_X^n(x^n) \\
&= \int_{\mathcal{A}^n} H\left(\Phi_{A(x^n)}[f_X]\right) dP_X^n(x^n) \\
&\geq \int_{\{x^n : B \subseteq A(x^n)\}} H\left(\Phi_{A(x^n)}[f_X]\right) dP_X^n(x^n) \\
&\geq H\left(\Phi_B[f_X]\right) \Pr\{B \subseteq A(X^n)\},
\end{aligned} \qquad (36)$$

where the last inequality follows from the monotonicity property in Claim 4 and $A(X^n)$ defined to be the set of distinct elements in $\{X_1, \ldots, X_n\}$. Now, for any $B \subseteq S$,

$$\begin{aligned}
\Pr(B \not\subseteq A(X^n)) &= \Pr\left(\bigcup_{b \in B} \{b \notin A(X^n)\}\right) \\
&\leq \sum_{b \in B} \Pr\{b \notin A(X^n)\} \\
&= \sum_{b \in B} (1 - \Pr\{X = b\})^n \\
&\leq |B| \left(1 - \min_{b \in B} \Pr\{X = b\}\right)^n \\
&\leq |B| \exp\left(-n \min_{b \in B} \Pr\{X = b\}\right).
\end{aligned} \qquad (37)$$

The proposition now follows by combining (36) with (37). $\square$

Besides being used in the proof of Theorem 5, Proposition 4 also gives the following corollary which will be used in the proof of Theorem 1.

**Corollary 6**
$$\liminf_{n \to \infty} H(Z_{n+1}|Z^n) \geq H\left(\Phi_S[f_X]\right),$$

*regardless of the finiteness of the right side of the inequality.*

*Proof of Corollary 6*:
Take a sequence $\{B_k\}$ of finite subsets $B_k \subseteq S$ satisfying

$$\lim_{k \to \infty} H\left(\Phi_{B_k}[f_X]\right) = H\left(\Phi_S[f_X]\right).$$

Proposition 4 implies, for each $k$,

$$\liminf_{n \to \infty} H(Z_{n+1}|Z^n) \geq H\left(\Phi_{B_k}[f_X]\right), \qquad (38)$$

completing the proof by taking $k \to \infty$ on the right side of (38). $\square$

*Proof of Theorem 5:*
Consider the case where $\{X_i\}$ are generated i.i.d. $\sim P$, where $P$ is a distribution on $\mathbb{N}$ and $p_j = P(X_1 = j)$ is a non-increasing sequence. Letting $D_l = \sum_{i=1}^{l} p_i \log \frac{1}{p_i}$ it follows by taking $B = B_l = \{1, \ldots, l\}$ in Proposition 4 that

$$H(Z_{n+1}|Z^n) \geq H(\Phi_{B_l}[f_X]) \left[1 - |B_l| \exp\left(-n \min_{b \in B_l} \Pr(X = b)\right)\right]$$
$$\geq D_l \left[1 - l \exp(-np_l)\right]$$

implying, by the arbitrariness of $l$,

$$H(Z_{n+1}|Z^n) \geq \max_l D_l \left[1 - l \exp(-np_l)\right]. \tag{39}$$

Consider now the distribution

$$p_i = P(X_1 = i) = \begin{cases} 0 & i = 1 \\ \frac{c(\varepsilon)}{i(\ln i)^{1+\varepsilon}} & i \geq 2 \end{cases} \tag{40}$$

for some $\varepsilon \in (0,1)$, where $c(\varepsilon)$ is the normalization constant. In this case

$$D_l = \sum_{i=2}^{l} \frac{c(\varepsilon)}{i(\ln i)^{1+\varepsilon}} \log \frac{i(\ln i)^{1+\varepsilon}}{c(\varepsilon)}$$
$$= \sum_{i=2}^{l} \frac{c(\varepsilon)}{i(\ln i)^{1+\varepsilon}} \left(\log i + (1+\varepsilon)\log(\ln i) - \log c(\varepsilon)\right)$$
$$= \sum_{i=2}^{l} \frac{c(\varepsilon)}{\ln 2} \left(\frac{\ln i}{i(\ln i)^{1+\varepsilon}}\right) + \sum_{i=2}^{l} \frac{c(\varepsilon)(1+\varepsilon)}{\ln 2} \left(\frac{\ln(\ln i)}{i(\ln i)^{1+\varepsilon}}\right) - \sum_{i=2}^{l} \frac{c(\varepsilon)}{\ln 2} \left(\frac{\ln c(\varepsilon)}{i(\ln i)^{1+\varepsilon}}\right)$$
$$= \sum_{i=2}^{l} \frac{c(\varepsilon)}{\ln 2} \left(\frac{1}{i(\ln i)^{\varepsilon}}\right) + \sum_{i=2}^{l} \frac{c(\varepsilon)(1+\varepsilon)}{\ln 2} \left(\frac{\ln(\ln i)}{i(\ln i)^{1+\varepsilon}}\right) - \sum_{i=2}^{l} \frac{c(\varepsilon)}{\ln 2} \left(\frac{\ln c(\varepsilon)}{i(\ln i)^{1+\varepsilon}}\right).$$

Observe that there exists $N_1' \in \mathbb{N}$ such that

$$D_l = \sum_{i=2}^{l} \frac{c(\varepsilon)}{\ln 2} \left(\frac{1}{i(\ln i)^{\varepsilon}}\right) + \sum_{i=2}^{l} \frac{c(\varepsilon)(1+\varepsilon)}{\ln 2} \left(\frac{\ln(\ln i)}{i(\ln i)^{1+\varepsilon}}\right) - \sum_{i=2}^{l} \frac{c(\varepsilon)}{\ln 2} \left(\frac{\ln c(\varepsilon)}{i(\ln i)^{1+\varepsilon}}\right)$$
$$> \sum_{i=2}^{l} \frac{c(\varepsilon)}{2\ln 2} \left(\frac{1}{i(\ln i)^{\varepsilon}}\right) \quad \forall l > N_1'$$
$$= \frac{c(\varepsilon)}{2\ln 2} \sum_{i=2}^{l} \frac{1}{i(\ln i)^{\varepsilon}} \quad \forall l > N_1'$$
$$> \frac{c(\varepsilon)}{2\ln 2} \int_{x=2}^{l+1} \frac{1}{x(\ln x)^{\varepsilon}} dx \quad \forall l > N_1'$$
$$= \frac{c(\varepsilon)}{2\ln 2} \left(\frac{(\ln(l+1))^{1-\varepsilon}}{1-\varepsilon} - \frac{(\ln(2))^{1-\varepsilon}}{1-\varepsilon}\right) \quad \forall l > N_1'.$$

Therefore there exists $N_2' > N_1'$ such that

$$D_l > \frac{c(\varepsilon)}{4\ln 2}(\ln(l+1))^{1-\varepsilon} \quad \forall l > N_2'. \tag{41}$$

Let $N_2 = (N_2'+1)^{(1+\varepsilon)/(1-\varepsilon)}$, $l_n = \lfloor n^{(1-\varepsilon)/(1+\varepsilon)}\rfloor$, and choose $N_3 > N_2$ such that

$$l_n = \lfloor n^{(1-\varepsilon)/(1+\varepsilon)}\rfloor \geq 1 \quad \forall n > N_3.$$

Combining (39) and (41) we get

$$\begin{aligned}H(Z_{n+1}|Z^n) \geq &D_{l_n}\left[1 - l_n \exp(-np_{l_n})\right] \quad \forall n > N_3\\ >& \frac{c(\varepsilon)}{4\ln 2}(\ln(l_n+1))^{1-\varepsilon}\left[1 - l_n \exp(-np_{l_n})\right] \quad \forall n > N_3\\ >& \frac{c(\varepsilon)}{4\ln 2}\left(\ln\left(\lfloor n^{(1-\varepsilon)/(1+\varepsilon)}\rfloor + 1\right)\right)^{1-\varepsilon}\left[1 - \lfloor n^{(1-\varepsilon)/(1+\varepsilon)}\rfloor \exp(-np_{l_n})\right] \quad \forall n > N_3\\ >& \frac{c(\varepsilon)}{4\ln 2}\left(\ln\left(n^{(1-\varepsilon)/(1+\varepsilon)}\right)\right)^{1-\varepsilon}\left[1 - n^{(1-\varepsilon)/(1+\varepsilon)} \exp(-np_{l_n})\right] \quad \forall n > N_3\\ >& \frac{c(\varepsilon)}{4\ln 2}\left(\frac{1-\varepsilon}{1+\varepsilon}\right)^{1-\varepsilon}(\ln n)^{1-\varepsilon}\left[1 - n\exp(-np_{l_n})\right] \quad \forall n > N_3 \tag{42}\end{aligned}$$

Since

$$p_i = P(X_1 = i) = \begin{cases} 0 & i = 1 \\ \frac{c(\varepsilon)}{i(\ln i)^{1+\varepsilon}} & i \geq 2 \end{cases}$$

there exists $N_4 > N_3$ such that

$$p_{l_n} > l_n^{-(1+\varepsilon)} \quad \forall n > N_4.$$

From (42) we get

$$\begin{aligned}H(Z_{n+1}|Z^n) >& \frac{c(\varepsilon)}{4\ln 2}\left(\frac{1-\varepsilon}{1+\varepsilon}\right)^{1-\varepsilon}(\ln n)^{1-\varepsilon}\left[1 - n\exp(-np_{l_n})\right] \quad \forall n > N_3\\ >& \frac{c(\varepsilon)}{4\ln 2}\left(\frac{1-\varepsilon}{1+\varepsilon}\right)^{1-\varepsilon}(\ln n)^{1-\varepsilon}\left[1 - n\exp\left(-nl_n^{-(1+\varepsilon)}\right)\right] \quad \forall n > N_4\\ >& \frac{c(\varepsilon)}{4\ln 2}\left(\frac{1-\varepsilon}{1+\varepsilon}\right)^{1-\varepsilon}(\ln n)^{1-\varepsilon}\left[1 - n\exp\left(-n\left(\lfloor n^{(1-\varepsilon)/(1+\varepsilon)}\rfloor\right)^{-(1+\varepsilon)}\right)\right] \quad \forall n > N_4\\ >& \frac{c(\varepsilon)}{4\ln 2}\left(\frac{1-\varepsilon}{1+\varepsilon}\right)^{1-\varepsilon}(\ln n)^{1-\varepsilon}\left[1 - n\exp\left(-n\left(n^{(1-\varepsilon)/(1+\varepsilon)}\right)^{-(1+\varepsilon)}\right)\right] \quad \forall n > N_4\\ >& \frac{c(\varepsilon)}{4\ln 2}\left(\frac{1-\varepsilon}{1+\varepsilon}\right)^{1-\varepsilon}(\ln n)^{1-\varepsilon}\left[1 - n\exp(-n^\varepsilon)\right] \quad \forall n > N_4. \tag{43}\end{aligned}$$

Finally, there exists $N_5 > N_4$ such that $N_5 \geq 2$ and

$$1 - n\exp(-n^\varepsilon) > 1 - \varepsilon \quad \forall n > N_5.$$

From (43) we get

$$\begin{aligned}
H(Z_{n+1}|Z^n) &> \frac{c(\varepsilon)}{4\ln 2}\left(\frac{1-\varepsilon}{1+\varepsilon}\right)^{1-\varepsilon}(\ln n)^{1-\varepsilon}\left[1-n\exp(-n^\varepsilon)\right] \quad \forall n > N_4 \\
&> \frac{c(\varepsilon)}{4\ln 2}\frac{(1-\varepsilon)^{2-\varepsilon}}{(1+\varepsilon)^{1-\varepsilon}}(\ln n)^{1-\varepsilon} \quad \forall n > N_5 \\
&= \frac{c(\varepsilon)}{4\ln 2}\frac{(1-\varepsilon)^{2-\varepsilon}}{(1+\varepsilon)^{1-\varepsilon}}(\ln 2 \log n)^{1-\varepsilon} \quad \forall n > N_5 \\
&= \frac{c(\varepsilon)}{4(\ln 2)^\varepsilon}\frac{(1-\varepsilon)^{2-\varepsilon}}{(1+\varepsilon)^{1-\varepsilon}}(\log n)^{1-\varepsilon} \quad \forall n > N_5 \\
&> \frac{c(\varepsilon)}{8(\ln 2)^\varepsilon}\frac{(1-\varepsilon)^{2-\varepsilon}}{(1+\varepsilon)^{1-\varepsilon}}(2\log n)^{1-\varepsilon} \quad \forall n > N_5 \\
&= \frac{c(\varepsilon)}{8(\ln 2)^\varepsilon}\frac{(1-\varepsilon)^{2-\varepsilon}}{(1+\varepsilon)^{1-\varepsilon}}\left(\log n^2\right)^{1-\varepsilon} \quad \forall n > N_5 \\
&> \frac{c(\varepsilon)}{8(\ln 2)^\varepsilon}\frac{(1-\varepsilon)^{2-\varepsilon}}{(1+\varepsilon)^{1-\varepsilon}}(\log(n+1))^{1-\varepsilon} \quad \forall n > N_5.
\end{aligned}$$

Thus (35) is satisfied under the distribution in (40) with any $\varepsilon \in (0, \min\{\delta, 1\})$. □

## 6 Conclusion

We have characterized the relationship between the entropy rate of a source and that of its pattern process for i.i.d., discrete Markov, discrete stationary ergodic, and a broad family of uncountable alphabet stationary ergodic processes. Besides determining the fundamental compression limits for a pattern sequence, the relationship between pattern and process entropy rate helps to quantify how much of the total information contained in the original stochastic process is encompassed in its pattern sequence. For the case where the pattern entropy rate is infinite, we characterized achievable growth rates for the block entropy of a pattern sequence.

## A  Proof of Theorem 1

If $|S| = 0$, then $\Pr\{\exists\, i \neq j : X_i = X_j\} = 0$. Therefore $H(Z^n) = 0$ for all $n$. This implies that $H(\mathbf{Z}) = 0$ which agrees with Theorem 1. Hence we just need to prove Theorem 1 for the case where $|S| > 0$.

Note that Corollary 6 and the fact that regardless of the finiteness of $H(X_1)$, $H(Z_1) < \infty$ and $H(Z_n|Z^{n-1}) < \infty$ for all $n$ gives

$$\liminf_{n\to\infty} \frac{H(Z^n)}{n} \geq H(\tilde{X}_1). \tag{44}$$

For the reverse inequality, look at

$$\limsup_{n\to\infty} \frac{H(Z^n)}{n} \overset{(a)}{\leq} \limsup_{n\to\infty} \frac{H(\tilde{X}^n)}{n}$$
$$\overset{(b)}{=} H(\tilde{X}_1), \qquad (45)$$

where $(a)$ comes from the fact that given $\tilde{X}^n$ we can reconstruct $Z^n$ w.p. 1 and $(b)$ from the fact that $\{\tilde{X}_i\}$ is an i.i.d. process. Combining (44) and (45) and noting that $\{\tilde{X}_i\}$ is an i.i.d. process completes the proof of Theorem 1. $\qquad\square$